\documentclass[vci-paper]{vciart}
\usepackage{graphicx}

\usepackage{natbib}

 \usepackage{epsfig}
 \usepackage{subfigure}

\usepackage{dcolumn}
\usepackage{bm}

\usepackage{amssymb}

\newcommand{\mrad}{\mathrm{mrad}}

\newcommand{\sigmatrack}{\sigma_\mathrm{track}}

\begin{document}

\begin{frontmatter}

\title{Detectors for particle identification}

 \author[ULFMF,JSI]{Peter Kri\v{z}an}
 \ead{peter.krizan@ijs.si}
 \address[ULFMF]{Faculty of Mathematics and Physics, University of Ljubljana, Slovenia}
 \address[JSI]{Jo\v{z}ef Stefan Institute, Ljubljana, Slovenia}

\begin{abstract}
The paper reviews recent progress in particle identification methods.
A survey of motivations and requirements for particle identification
in various experimental environments is followed by the main emphasis, which
 is on the recent 
development of \v Cerenkov counters,  from upgrades of existing devices to 
a novel focusing radiator concept and new photon detectors. 
The impact of including a precise measurement of the 
time of arrival of \v Cerenkov photons to increase the kinematical region over which 
particle identification can be performed  is discussed. The progress 
in  dedicated time-of-flight counters with recently developed very fast single
photon detectors is also evaluated.
\end{abstract}


\end{frontmatter}

\section{Introduction}
\label{intro}
One of the main driving forces of the research and development of 
particle identification methods in the last decade  was 
the need to have excellent hadron identification for precision $B$ physics
measurements. For a statistically significant measurement of
CP violation in the $B$ system, the  tagging of  $B$ meson flavour with the kaon 
charge was an indispensable method. In  addition, for the study of rare 
few body hadronic decays of $B$ mesons it is essential to separate 
pions from kaons up to the kinematic limits of the experiment. Reliable hadron  
identification is also important in the search for quark-gluon plasma and in
studies of the nucleon structure.

Hadrons are identified by their mass, which is, in turn, determined by combining
the measurements of momentum and velocity. Assuming that momentum is inferred
from the measured radius of curvature in magnetic field, the remaining issue is to 
measure the velocity with a sufficient precision. This is either achieved by 
measuring the time-of-flight, ionization losses or \v Cerenkov angle of the particle.
In the present contribution  recent progress in  \v Cerenkov counters
and time-of-flight measurements is discussed.

The structure of the paper is as follows. We first review the 
 \v Cerenkov counters in some of the running or recently completed 
experiments. A number of new methods is presented, 
from upgrades of existing devices to a novel focusing radiator concept and 
new photon detectors. We then discuss
the benefits of including a precise measurement of the 
time of arrival of \v Cerenkov photons in order to reduce the dispersion error 
and the even more ambitious use   in a combined time-of-flight (TOF) and     
time-of-propagation (TOP)  counter.  Finally, we will discuss the progress 
in  dedicated TOF counters with recently developed very fast photon detectors.

\section{\v Cerenkov counters}
\label{present}

\subsection{DIRC at BaBar}
\label{present:BaBar}

\begin{figure}
\begin{center}
\resizebox{0.5\columnwidth}{!}{%
  \includegraphics{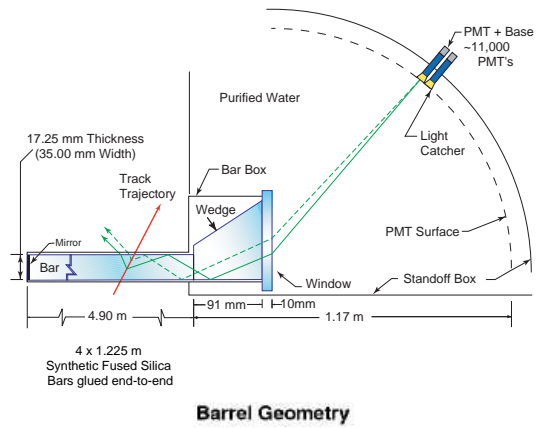}}
\resizebox{0.5\columnwidth}{!}{%
  \includegraphics{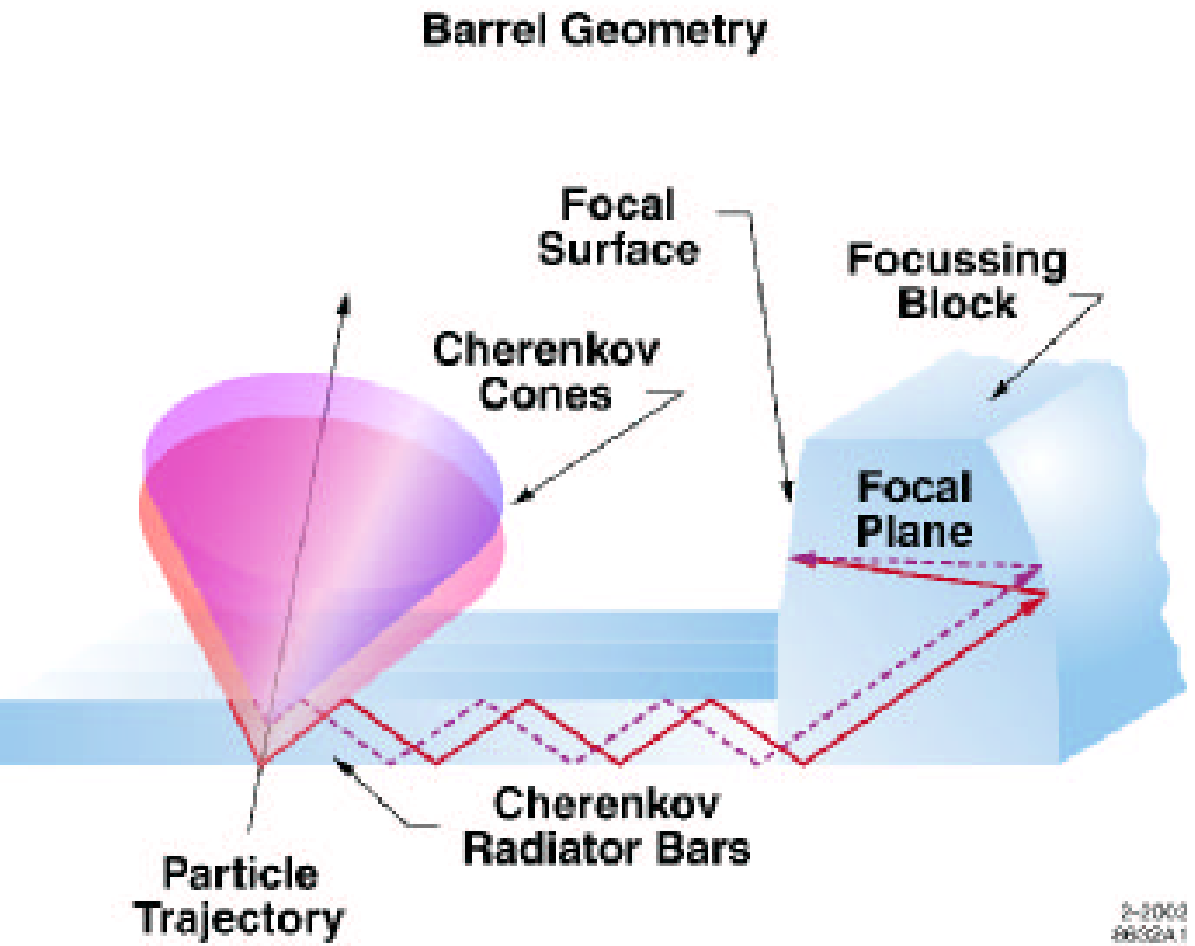}}
\end{center}
\caption{Principle of the DIRC counter. (top) and its upgrade, the focusing DIRC 
(bottom).\label{fig:dirc-princ}}
\end{figure}

The DIRC (Detector of Internally Reflected Cherenkov light) 
of the   BaBar spectrometer is a  special type of
a  ring-imaging \v Cerenkov counter \cite{dirc2}, based on the detection of 
\v Cerenkov photons trapped in the quartz radiator bar 
(Fig.~\ref{fig:dirc-princ}). The patterns 
on the photon detector are quite complicated, 
but result in well resolved peaks in the \v Cerenkov angle
distribution. The time of arrival of photons is used to
eliminate background from conversion events in the water tank,
to assign photons to proper tracks, and to
eliminate most of the ambiguities in the photon-track reconstruction.

\begin{figure}
\begin{center}
\resizebox{0.7\columnwidth}{!}{%
  \includegraphics{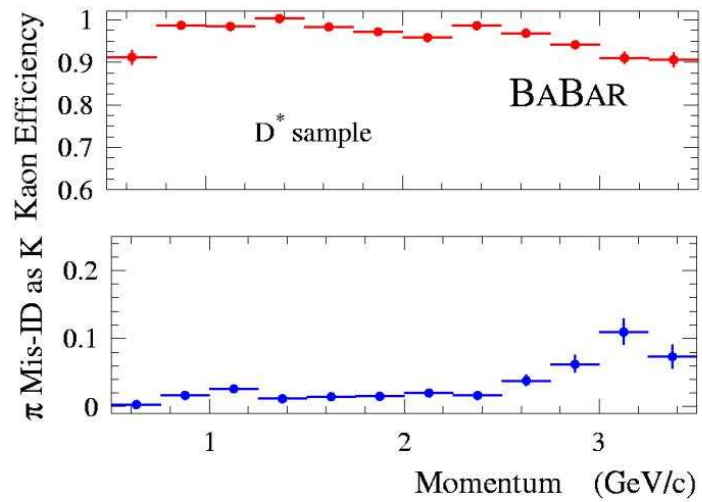}}
\caption{Particle identification with DIRC: $K$ efficiency and
 $\pi \to K$ misidentification probability.\label{fig:dirc-perf}}
\end{center}

\end{figure}

The basic performance parameters   of the counter are in excellent agreement with 
expectations  \cite{dirc2}. The single photon resolution amounts to 9.6~mrad.  
The number of photons  depends on the polar angle of 
the charged  track, but always stays above 20 for $\beta=1$ particles. 
The efficiency for kaon identification
exceeds 90\% in the momentum range 0.5~GeV/$c$ - 3~GeV/$c$, while the
probability that a pion is identified as a kaon stays at a few percent level
(Fig.~\ref{fig:dirc-perf}).

\subsection{ACC at Belle}
\label{present:Belle}

\begin{figure}
\begin{center}
\resizebox{0.9\columnwidth}{!}{%
  \includegraphics{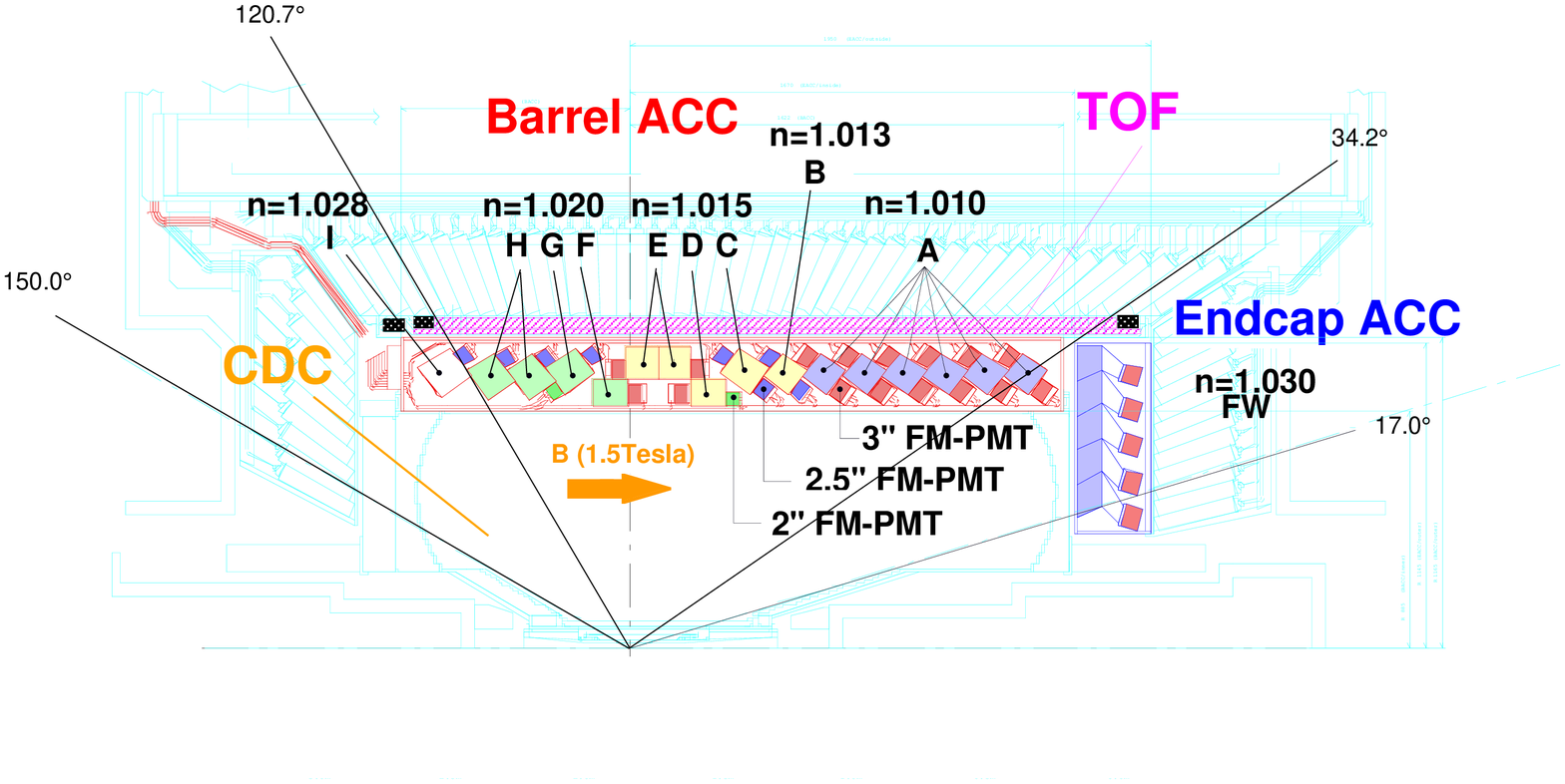}}
\end{center}
\caption{Aerogel \v Cerenkov Counter (ACC) of the Belle spectrometer.\label{fig:acc}}
\end{figure}

In the  Belle spectrometer the  separation of kaons from pions 
is performed with the Aerogel Cherenkov Counter (ACC),
 a threshold \v Cerenkov detector (Fig.~\ref{fig:acc})
with aerogel as radiator \cite{belle-acc}. The refractive index of the radiator 
is chosen so that pions emit \v Cerenkov light, while kaons stay below threshold.
Since the momentum spectrum  of particles becomes harder in the
forward direction, the refractive index of the modules gradually decreases from 
$n=1.028$ to $n=1.01$.
Note that while in the central (barrel) part it is possible to measure 
both the tagging kaons and the $B\to \pi\pi, K\pi$ decay products,
 only the former can be identified in the forward (end-cap) direction.
The kaon identification efficiency amounts to 90\% with the pion
fake probability  equal to 6\%   \cite{toru99}.

\subsection{ HERA-B RICH}
\label{herab}

\begin{figure}[htbp]
\centerline{\epsfig{file=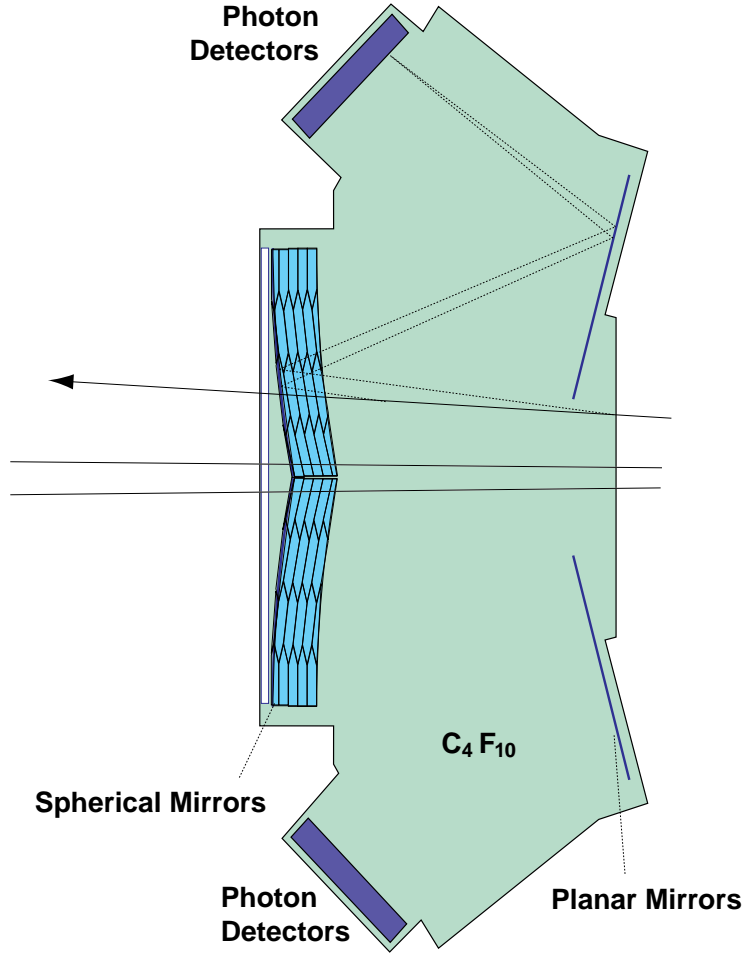,width=0.7\columnwidth,clip=}}
\caption[kk]{Layout of the HERA-B RICH counter.\label{fig:hbrich}}
\end{figure}

\begin{figure}[htbp]
\centerline{\epsfig{file=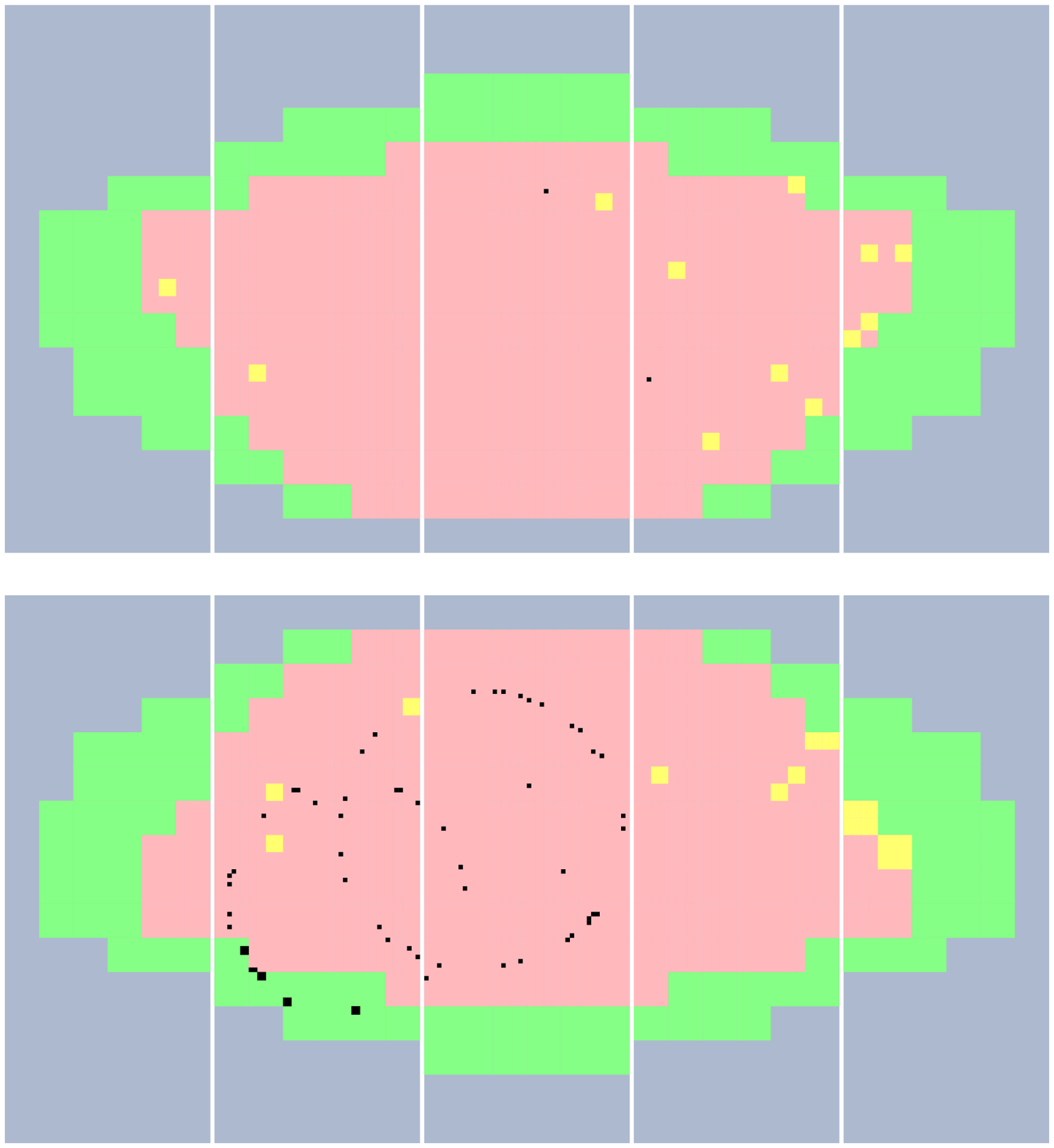,width=0.4\columnwidth,angle=-90,clip=}}
\centerline{\epsfig{file=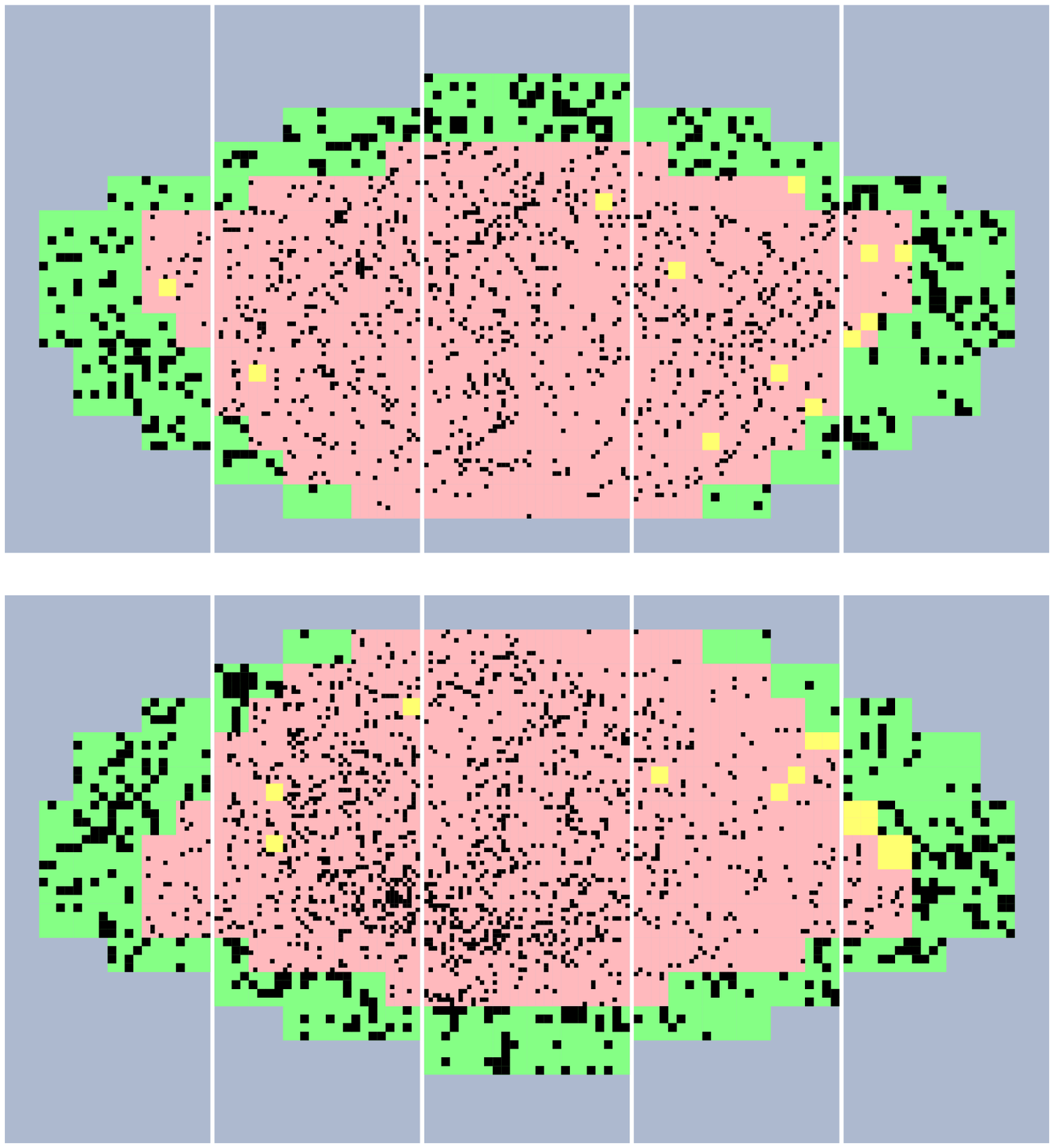,width=0.4\columnwidth,angle=-90,clip=}}
\caption[kk]{Events recorded in the lower half of the HERA-B RICH: 
a low multiplicity event (top) with  two clear rings, and a typical event (bottom).
\label{fig:hb_bckg}}
\end{figure}

The HERA-B collaboration 
was the first to employ multianode photomultiplier tubes in a RICH 
counter (Fig.~\ref{fig:hbrich})  \cite{hbrich-main}, after having shown 
that the Hamamatsu R5900 PMTs  (versions M16 and M4 with 16 and 
4 square shaped channels, respectively) have a  
capability to detect single photons with high efficiency,
and little cross-talk  \cite{pmt-first}.
The photon detector performed very well, 
showing  clear rings with very few noisy channels ($<0.5$\%) 
 even in a very hostile environment of a hadron machine,
as can be seen from Fig.~\ref{fig:hb_bckg}. 
\begin{figure}[bht]
\centerline{\epsfig{file=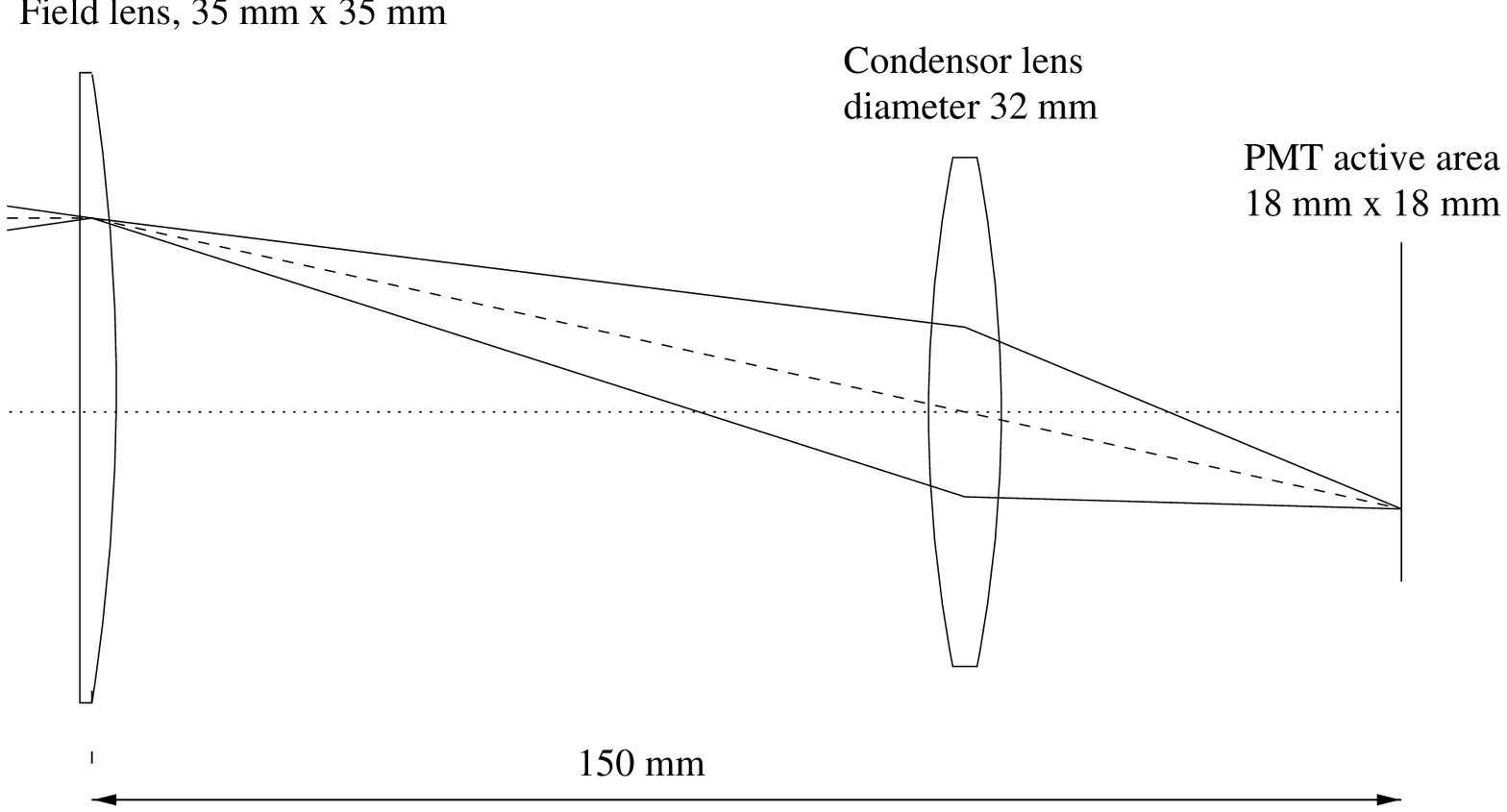,width=0.35\columnwidth,angle=-90}}
\caption[kk]{ The optical system for light collection and demagnification.
The two rays shown in full line correspond to photons with incident angles of
$\pm 100$~mrad
with respect to the normal incidence (dashed line).\label{fig:lens_syst}}
\end{figure}

A  common drawback of vacuum based photon detectors is a rather large fraction of 
dead area. While for single channel PMTs reflective cones can be used, 
multichannel PMTs need an imaging system.  In the HERA-B RICH a system 
of two lenses, shown in Fig.~\ref{fig:lens_syst},
 was used to demagnify the image on the focal surface by a factor of 2
\cite{dan-lyon}. The 
system of lenses is used both to reduce the dead area, as well 
as to adapt the required 
granularity of the photon detector to the granularity of the multianode PMT.     
The transparency of the system is high in the region of high quantum efficiency of the tube. 
It also has a flat acceptance for photons with incidence angles below 140~mrad 
as required by the detector geometry. 

\begin{figure}[htbp]
\centerline{\epsfig{file=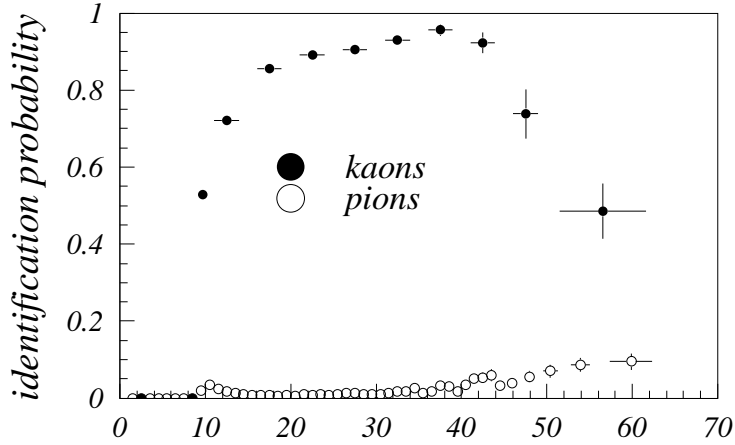,width=0.7\columnwidth,angle=0}}
\caption[kk]{Particle identification in the HERA-B RICH: efficiencies to
identify a kaon and misidentify a pion, as deduced from the $\phi \to K^- K^+$ and
$K_S \to \pi^- \pi^+$ decays.\label{fig:jpsieff}}
\end{figure}

The experience of the HERA-B collaboration shows that a RICH counter
can safely be operated even at high track densities with counting
 rates exceeding 1~MHz per channel in the hottest part of the photon detector 
\cite{hbrich-main}.
 A typical event is shown in the lower plot of Fig.~\ref{fig:hb_bckg}.
No  degradation of performance has been observed over the five years of operation.
The relevant RICH parameters were determined from the data.  On average 
32 photons were detected per $\beta=1$ ring.
The single photon  resolution was 0.7~mrad in the finer granularity area 
covered by R5900-M16 PMTs. All parameters are in very good agreement 
with expectations. The particle identification capabilities are illustrated in 
Fig.~\ref{fig:jpsieff}.
 
\subsection{COMPASS, Hades and ALICE RICH counters}
\label{csi}

The  RICH counters of COMPASS \cite{compass-rich}, Hades \cite{hades} and 
ALICE\cite{alice-rich} experiments  employ
 multiwire chambers with a solid CsI layer, evaporated onto the cathode pads, 
as the photon detector.
The material requires a high purity chamber gas, usually  methane with water and 
oxygen content of order ppm, as well as careful handling of the photo-cathodes. 
A lot of R\&D (influence of substrate, thermal conditioning, thickness,
evaporation procedure, transfer to the chamber) was needed to make the technique mature
\cite{csi95,alice-csi}. A special evaporation plant with a sophisticated
conditioning facilities and quality assessment controls was constructed at CERN
for this purpose.
While Compass and Hades RICH counters have been running stably for several years, the 
ALICE RICH is under construction  \cite{alice-vci07}.

In the Compass experiment the central region of the RICH counter was recently 
upgraded to enable running at higher beam intensities, higher trigger rates and
with a nanosecond time resolution for background suppression
\cite{compass-mapmt}. The wire chamber based photon detector was replaced 
by an array of multianode PMTs very similar to the HERA-B RICH (Sec. \ref{herab}). 
The PMTs, again Hamamatsu 16 channel tubes, have quartz windows, and the 7:1 
demagnification two-lens light collection system is made of  quartz lenses. This
results in a larger number of detected photons ($\approx$60 per ring for $\beta = 1$ 
particles).
The measured \v Cerenkov angle resolution per track is ~$\approx$0.3~mrad, allowing a  
2$\sigma$ $\pi/K$ separation at 60~GeV/$c$. The kaon identification efficiency 
exceeds 90\% with $\pi-K$ misidentification probability at $\approx$1\%.

\subsection{LHCb RICH}
\label{lhcb}

\begin{figure}[htbp]
\centerline{\epsfig{file=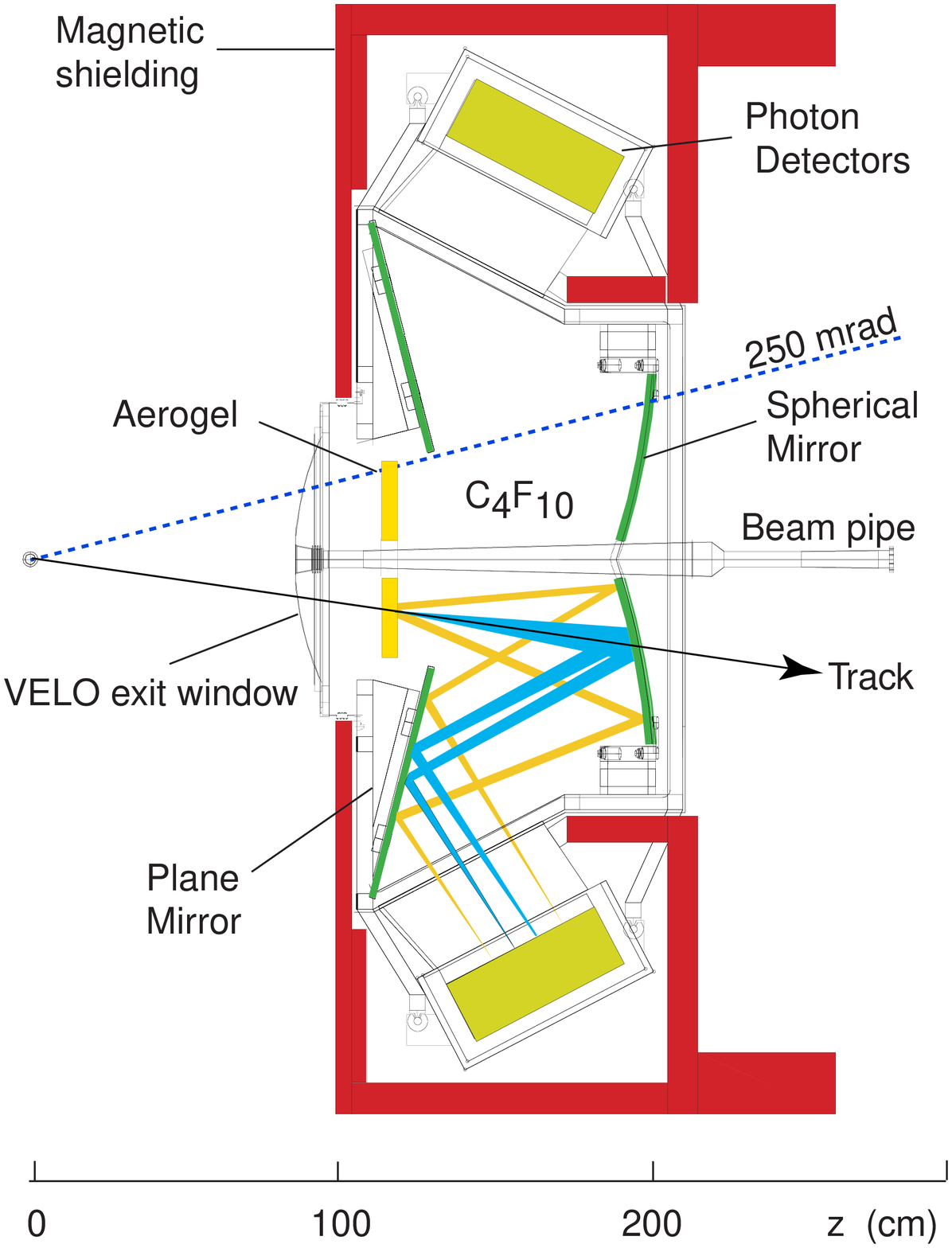,clip=,angle=0,width=0.5\columnwidth}
\epsfig{file=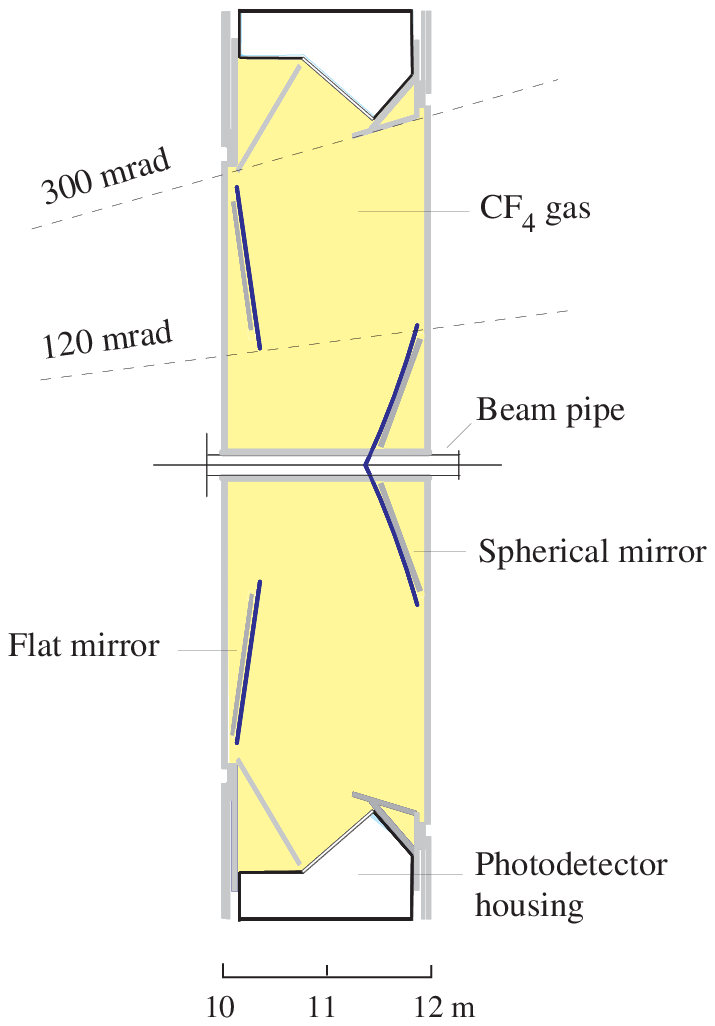,clip=,angle=0,width=0.5\columnwidth}}
\caption[kk]{RICH counters for the LHCb experiment.\label{fig:lhcb-riches}}
\end{figure}

In the next generation of $B$ physics experiments at hadron colliders  
the identification of hadrons will  again be essential for both the tagging of 
$b$-flavour in
$CP$ violation and for mixing measurements, as well as for the identification of 
particles in hadronic final states such as $B\to \pi^+\pi^-$.
In particular this is so because more channels contribute to the background than in the 
case of experiments at $\Upsilon (4s)$ (e.g. $B_s\to K^+K^-$). 
In the LHCb experiment the RICH counter is designed to cover $\pi /K$ separation 
between 1 GeV/c and 150 GeV/c \cite{lhcb-tdr}.
The kinematic region covered by a RICH counter,  from $p_{min}$ to $p_{max}$,
depends on the threshold momentum for the lighter of the two particles one wants to 
separate,  and on the resolution in 
\v Cerenkov angle (ultimately given by the dispersion in the radiator medium).
It turns out that for most radiators $p_{max}/p_{min} \approx 4-7$ \cite{rich-limits}.
For the large
 kinematic interval in the LHCb experiment one would therefore require three radiators. They are
 arranged  in two counters \cite{lhcb-tdr} as shown in Fig.~\ref{fig:lhcb-riches}.

\begin{figure}[htbp]
\centerline{\epsfig{file=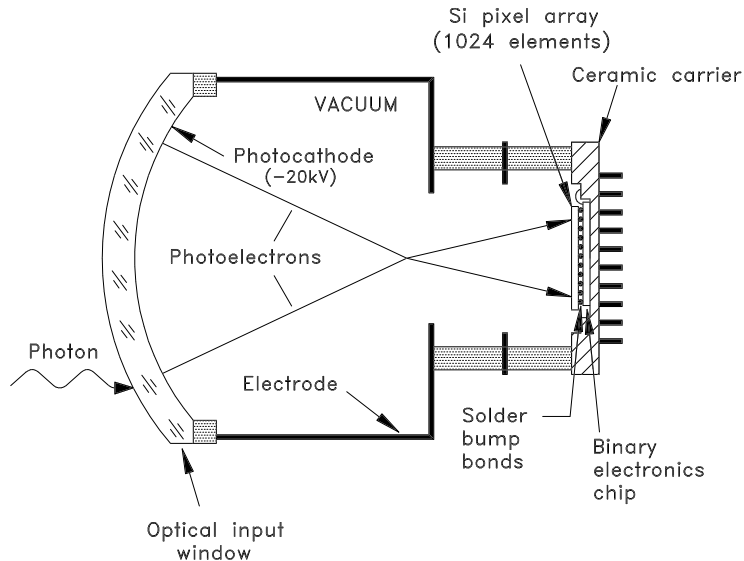,clip=,angle=0,width=0.7\columnwidth}}
\caption[kk]{Hybrid photon detector (HPD) of the LHCb experiment.  
The outer diameter of the detector is 17~cm.\label{fig:lhcb-hpd}}
\end{figure}

Several photon detectors were considered as candidates for 
the system of RICH counters;  
a photon detector based on  multianode PMTs with a single quartz lens light collection system
\cite{lhcb-mapmt},
as well as two types of  hybrid photon detectors (HPD). In the latter,  
 a vacuum photosensitive device is combined with charged
 particle detection in a silicon detector with pixel readout \cite{lhcb-hpd}. 
As  shown in Fig.~\ref{fig:lhcb-hpd}, 
photoelectrons are accelerated by the electric field across a potential difference of 
about 20 kV towards a silicon detector with integrated electronics. 
The results of beam tests of the final module, developed in collaboration with the
DEP company,  are very encouraging \cite{vidal-siltjes}. 

\section{Recent developments in RICH counters}
\label{upgrades}

For the next round of $B$ physics experiments at upgraded 
 $e^+e^-$ machines ('Super $B$ factories') with luminosities
exceeding $10^{35}$~cm$^{-2}$s$^{-1}$, considerable improvements of particle 
identification
devices  are envisaged in order  to cope with higher rates and with more stringent
 requirements on separation capabilities for rare decay channels.

\subsection{Upgrades at Belle}
\label{upgrades:Belle}
Two systems  are being  considered for the upgrade of the  Belle spectrometer
  \cite{bib:superkekb-loi}. 
For the  barrel region a time-of-propagation (TOP) counter  is being studied. This is   
a kind of DIRC counter in which the \v Cerenkov
angle is deduced by measuring one of the coordinates and 
the time of arrival of  \v Cerenkov photons  with high precision \cite{Ohshima}.
Multianode micro-channel plate (MCP) PMTs, including a detector 
with a GaAsP photocathode, are being investigated  \cite{tof}.

\begin{figure}[hbt]
\begin{center}
\resizebox{0.35\columnwidth}{!}{\includegraphics{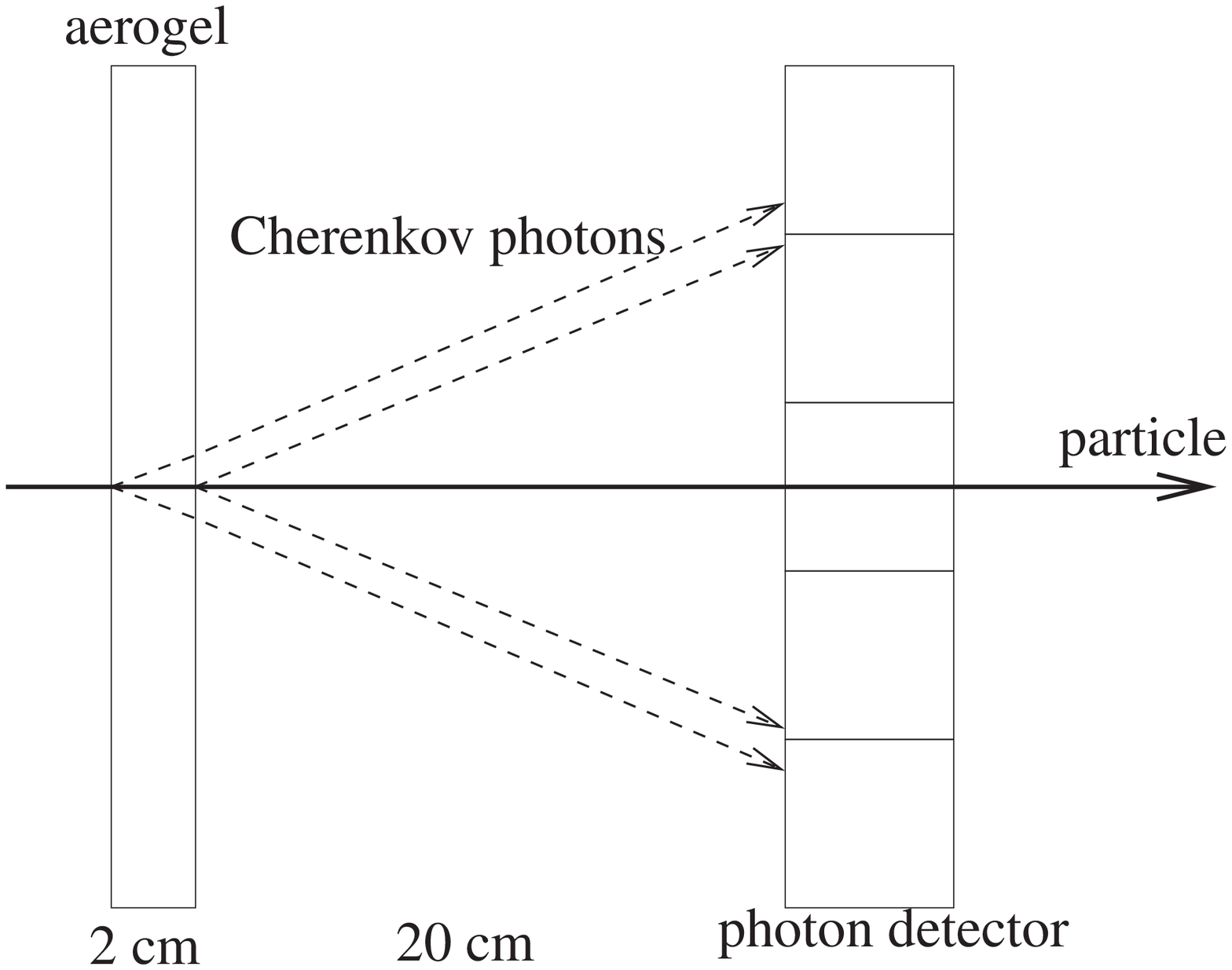}}
\resizebox{0.55\columnwidth}{!}{\includegraphics{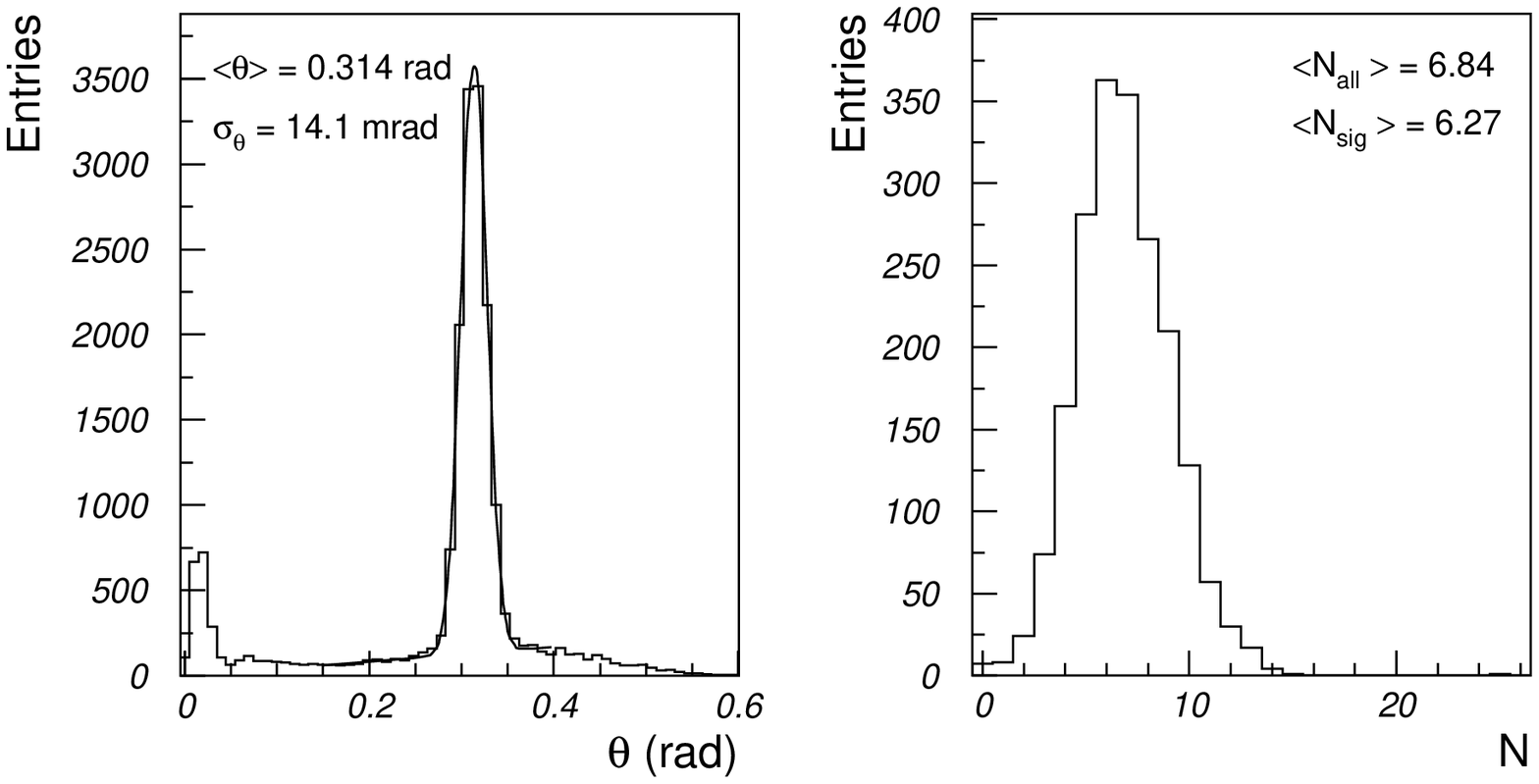}}
\end{center}
\caption[ll]{Proximity focusing RICH for the Belle upgrade  in the end-cap region: 
schematic (a), measured single photon resolution (b) and the number of 
detected photons (c) for a 2~cm thick aerogel tile.\label{fig:proxfoc}}
\end{figure}

For the end-cap region,  a proximity focusing RICH with aerogel as radiator 
is being tested (Fig.~\ref{fig:proxfoc}). The counter should have a low threshold
for pions, and should enable good separation of pions and kaons up to  4~GeV/$c$.
Another benefit of such a counter 
 would also be a reasonable $e/\mu/\pi$ separation at low momenta, which is of importance for 
the studies of rare $B \to K \ell \ell$ decays.

 The feasibility of such a counter was studied in a series of beam tests
\cite{2ndBeamTest,pk-rich04}. Hamamatsu H8500  ('flat panel') PMTs
were used as  photon detectors \cite{H8500data}. As can be seen from 
Fig.~\ref{fig:proxfoc}, the \v Cerenkov peak is well pronounced above a small background,
mainly coming from \v Cerenkov photons which were Rayleigh scattered in the radiator.
The resolution in the \v Cerenkov angle measurement ($\sigma_{\theta}=14$~mrad) and 
the number of detected photons
($N \approx 6$, Fig.~\ref{fig:proxfoc}) agree well with expectations. The resulting resolution 
per track is about 5.7~mrad.  Since the difference 
in \v Cerenkov angle of pions and kaons is 22~mrad at  4~GeV/$c$, such
 a counter would allow a good (about 4$\sigma$) 
separation up to  the kinematic limit of the experiment.

\begin{figure}
 \centerline{\epsfig{file=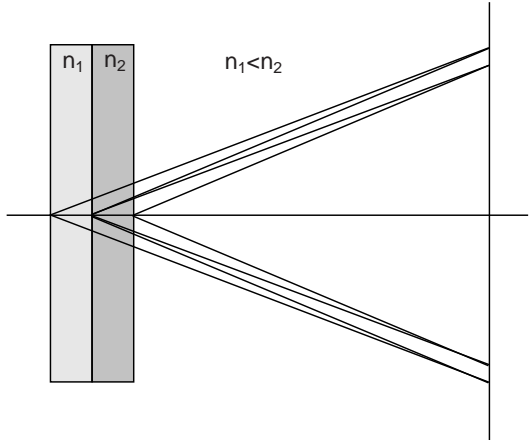,clip=,angle=0,width=0.5\columnwidth}}
 \caption{Proximity focusing RICH with  a nonhomogeneous aerogel radiator 
          in the  focusing configuration. \label{princ}}
\end{figure}

The key issue in the performance of a proximity focusing RICH counter is to
improve the \v Cerenkov angle resolution per track $\sigmatrack = \sigma_{\theta}/\sqrt{N}$.
As it turns out,  the optimal thickness is expected to be around $20~\mathrm{mm}$ 
\cite{2ndBeamTest,nim-multi}.
However, this limitation can be overcome in a proximity focusing RICH 
with a non-homogeneous radiator~\cite{nim-multi,pk-hawaii,danilyuk,multi-rad-ana}.  
By judiciously choosing the
refractive indices of consecutive aerogel radiator layers, one may achieve overlapping
of the corresponding \v Cerenkov rings on the photon detector (Fig.~\ref{princ}) 
\cite{multi-rad-ana}. 
This represents a sort of focusing
of the photons within the radiator, and eliminates or at least considerably reduces the spread
due to emission point uncertainty.  Note that such a tuning of  
 refractive indices for individual layers is only possible with aerogel, 
which may be produced with any desired refractive index in the range 1.01-1.07 
\cite{aerogel-new}. 
The dual radiator combination can readily be extended to
more than two aerogel radiators. In this case, the indices of aerogel layers should
gradually increase from the upstream to the downstream layer.

\begin{figure}
 \centerline{\epsfig{file=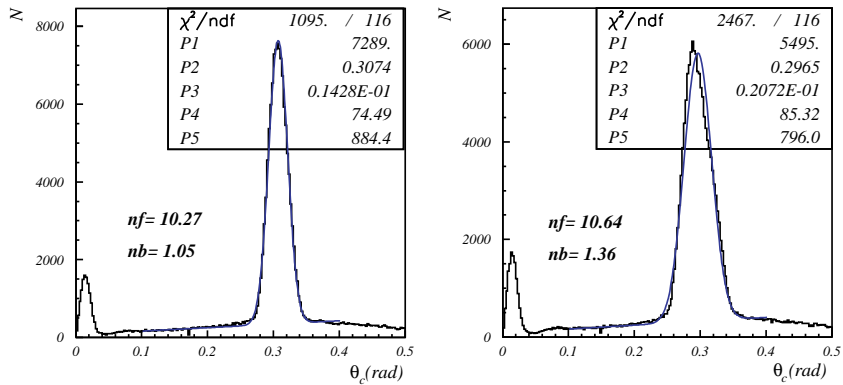,clip=,angle=0,width=0.8\columnwidth}}
 \caption{The accumulated distribution of \v Cerenkov photon hits depending
          on the corresponding \v Cerenkov angle  for a 4cm homogeneous radiator
          (right) and a focusing configuration with $n_{1}$~=~1.046,
          $n_{2}$~=~1.056 (left).  \label{foc}}
\end{figure}

In Fig.~\ref{foc}, we compare the data for two 4~cm thick radiators;
one with
aerogel tiles of equal refractive index (n = 1.046), the other with the
focusing arrangement ($n_{1} = 1.046 , n_{2} = 1.056$). The
improvement is clearly visible.
The single photon resolution $\sigma_\theta=14.3~\mrad$ for the dual radiator 
is considerably smaller than the corresponding value for the single refractive
index radiator 
($\sigma_\theta=20.7~\mrad$), while the number of detected photons is the same in 
both cases.

The present R\&D efforts are oriented mainly toward the development of a photon 
detector which works in the high magnetic field (1.5~T) of the Belle spectrometer.
The Burle 85011 MCP PMT  \cite{burle}, a multichannel device with 8x8 channels 
and 6~mm $\times$ 6~mm large pads,  has been shown to perform very well as a 
 detector of \v Cerenkov photons in a RICH counter \cite{mcp}.

Two further detectors are being studied.
In a joint effort with the Hamamatsu company a new type of hybrid photon detector of the
proximity focusing type is being developed. As the second option, silicon photomultipliers,
 novel semiconductor photon detectors are being tested \cite{icfa,rusi,renker}.
 They have several advantages: insensitivity to the high magnetic fields, 
lower operation voltage, and less material in comparison to the conventional photomultiplier 
tubes.  They also have high peak photon detection efficiency ($approx$ 20\%), 
high gain of $\approx 10^6$ and good time response. Due to their dimensions, they allow compact, 
light and robust mechanical designs. 
All this would make them a very promising candidate for a detector of \v Cerenkov photons 
in a RICH counter. However, due to the  serious disadvantage of a very high dark rate 
($\approx 10^6$~Hz/mm$^2$), they have up to now never been used in ring imaging 
\v Cerenkov detectors, where single photon detection is required at low noise. 
Because the \v Cerenkov light is prompt,   this problem could in principle  be reduced 
by using a narrow time window ($<10$~ns) for signal
collection. In addition, it is  possible to 
further increase the signal-to-noise ratio by using light collection systems \cite{rok-vci07}.

\subsection{Focusing DIRC}
\label{upgrades:BaBar}

As  an upgrade of the DIRC counter, a layout with a considerably smaller stand-off box
is being considered \cite{dirc3} as shown in  Fig.~\ref{fig:dirc-princ}. 
Such a change would significantly reduce the beam related 
background. However, in order not to degrade the angular resolution, single channel 
PMTs would have to be replaced by multichannel devices among which the
Hamamatsu flat panel 64 channel PMTs and the Burle 64 channel micro-channel plate (MCP) 
PMTs have been studied.
Two further changes are considered which should considerably improve the angular 
resolution.
The uncertainty in the emission point along the track will be eliminated by
focusing optics of the expansion volume made of quartz. The chromatic error will
be reduced by measuring the time of arrival of \v Cerenkov photons with a resolution
of  50-100~ps, from which the wavelength of each photon can be estimated.   The 
expected resulting total angular error is 4-5mrad per single photon, which makes 
it possible to achieve an angular resolution of 1.5~mrad per track in principle.
While the present BaBar DIRC achieves a $2.7\sigma$ $\pi/K$ separation at 4~GeV/$c$, the 
equivalent performance of the upgraded DIRC would be  $4.3\sigma$ at 4~GeV/$c$ 
for photons traveling a full bar length of 3-4~m.

To demonstrate the capability of eliminating
the chromatic error contribution, tests of candidate photon
detectors were carried out to evaluate their timing resolution and homogeneity of 
response \cite{vavra-elba}.
By using a pulsed  laser diode and dedicated read-out electronics, the time resolution
of the MCP PMTs was measured. An excellent time resolution of 30~ps was achieved
\cite{vavra-30ps,vavra-vci07}.  
The performance of a focusing DIRC prototype was successfully tested in 
a test beam \cite{vavra-vci07}.

\section{Recent progress in time-of-flight counters}
\label{sec:tof}

\begin{figure}
\centerline{\includegraphics[width=0.7\columnwidth]{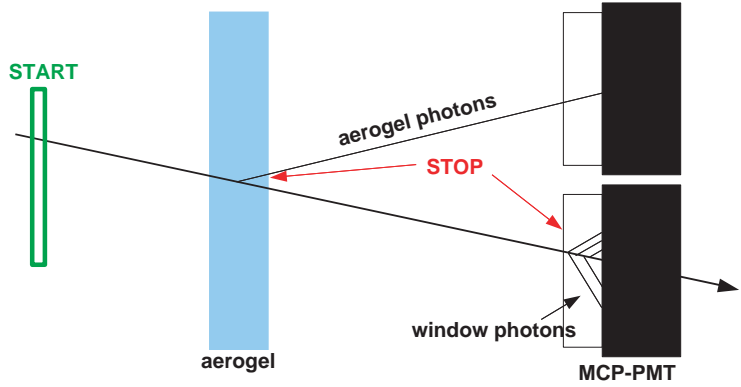}}
\caption{Schematic view of a combined RICH and TOF counter.\label{richtof}}
\end{figure}

A proximity focusing RICH counter such as the one discussed in section \ref{upgrades:Belle}
is also a very fast counter, in 
particular if a micro-channel plate (MCP) PMT is used as the photon detector. 
With its excellent timing properties, the same device could also serve  
as a time-of-flight counter and thus supplement other identification methods,
in particular for low momentum tracks. \v Cerenkov photons emitted in the 
radiator medium as well as in the entrance window of the PMT can be used 
for the time-of-flight measurement (Fig.~\ref{richtof}).
This allows a positive identification
also of particles that would be below the \v Cerenkov threshold in the aerogel
radiator (kaons and protons in the region around 1~GeV). Consequently, a good 
separation of kaons and protons would be possible in this region as well.

\begin{figure}
\centerline{\includegraphics[width=0.4\columnwidth]{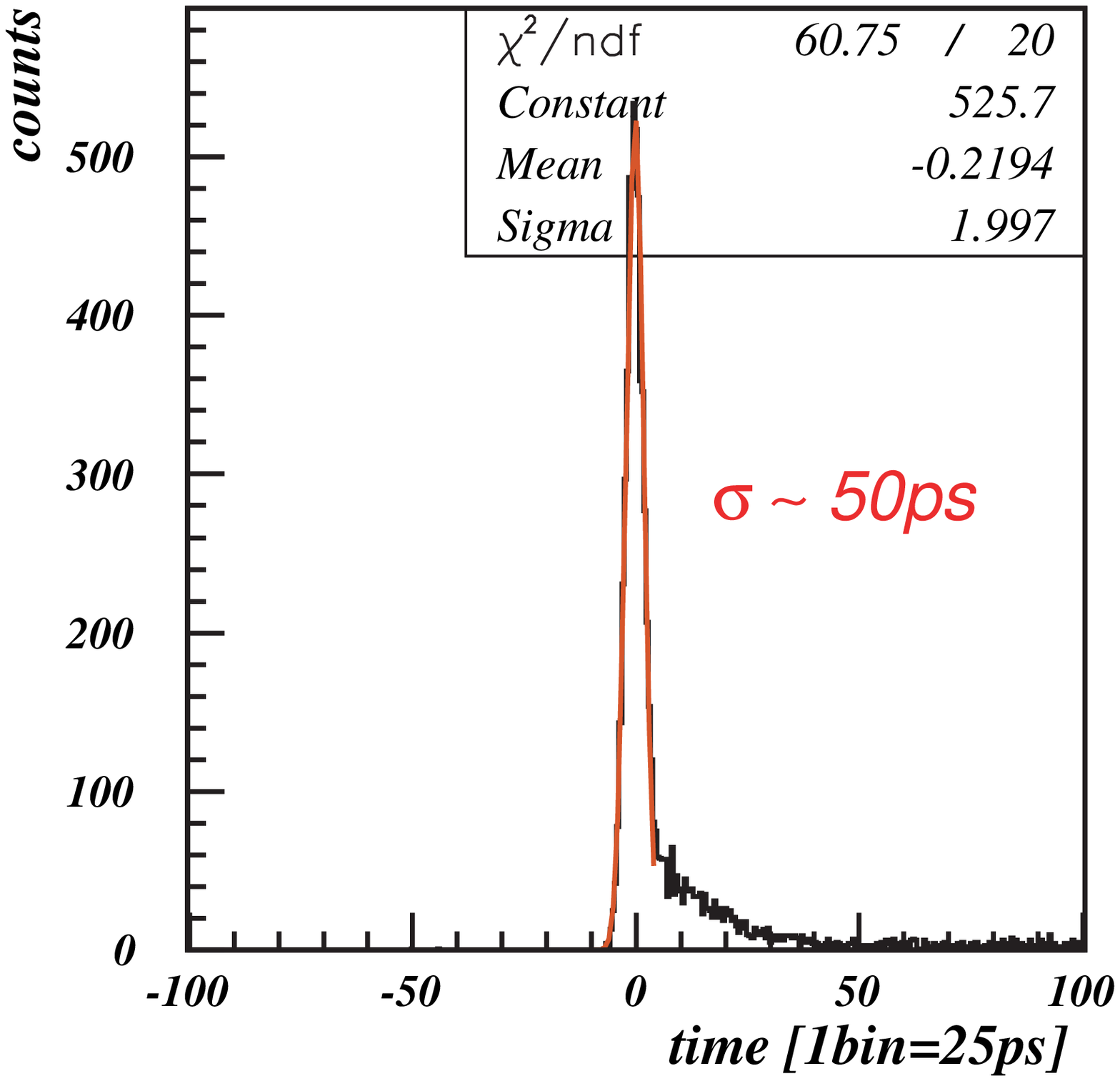}
\includegraphics[width=0.4\columnwidth]{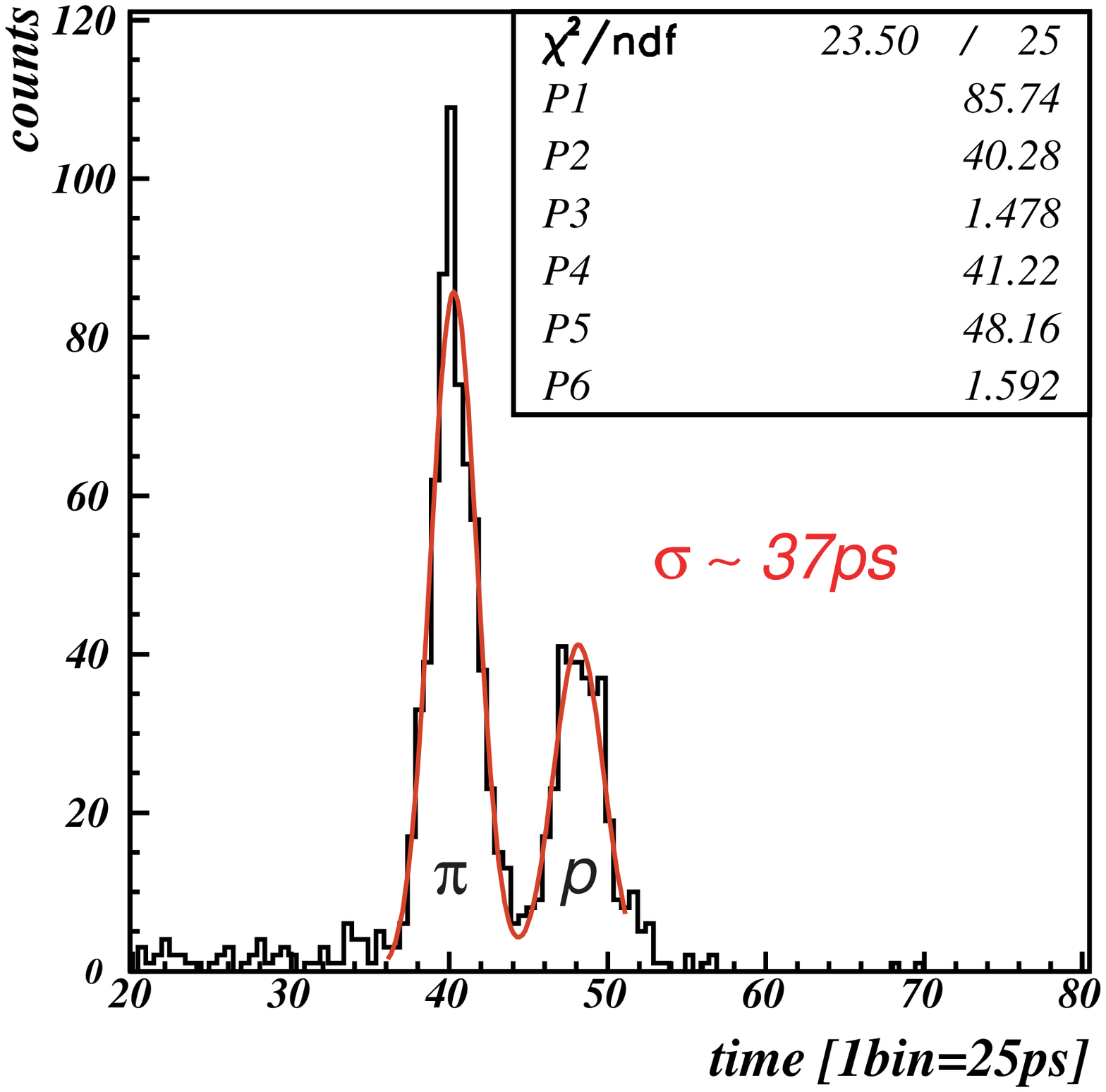}}
\caption{The distribution of the time of flight as measured by single 
\v Cerenkov photons from the  aerogel radiator (left), and by  \v Cerenkov 
photons from the PMT window, shown for 2~GeV/c pions and protons (right).
\label{ringtdc}}
\end{figure}

To test this idea, in a beam test both the hit position and its time were 
registered with a Burle 85011 MCP PMT \cite{rich-tof}. The distributions of hits,
depending on their time of flight for \v Cerenkov photons from the aerogel, 
is plotted in Fig.~\ref{ringtdc}.  Fitting this distribution
with a Gaussian function yields a  standard deviation of about 50~ps.
For 10 detected hits per track this would correspond to a  time-of-flight 
resolution of 20~ps.

An excellent resolution is also found in the  time-of-flight distribution 
as determined by \v Cerenkov photons from the  PMT window. As can be 
seen in Fig.~\ref{ringtdc}, the standard deviation of the distribution
for pions is 37~ps. We can also see that pions are clearly separated from
protons even in this very compact set-up with a very short flight path of 65~cm.
In the Belle spectrometer, where the typical flight path is about 2~m, the measured 
performance would correspond to a 6~$\sigma$ separation between
pions and kaons at 2~GeV/c, and a 3.5~$\sigma$ separation between
protons and kaons at 4~GeV/c.

\begin{figure}
 \centerline{\epsfig{file=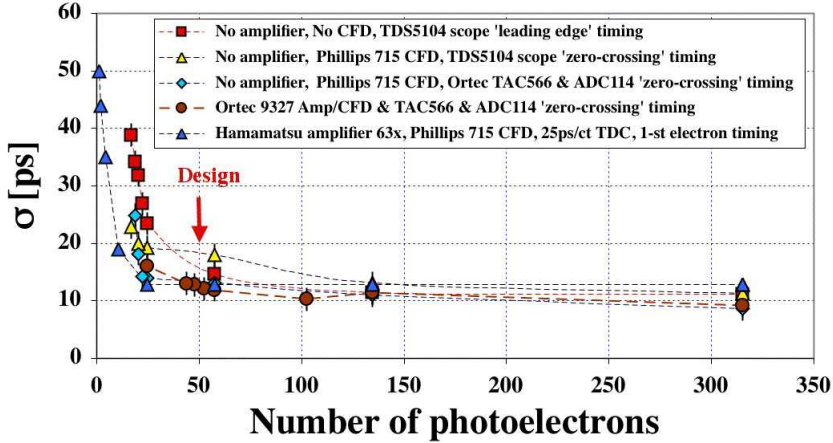,clip=,angle=0,width=0.8\columnwidth}}
 \caption{Timing resolution as a function of number of
photoelectrons  using various methods for a MCP-PMT with 10~$\mu$m hole diameter 
\cite{vavra-vci07}.\label{fig:tres-vavra}}
\end{figure}

To further improve the time resolution, one could increase the number of photons 
to about 50 by adding an additional  1~cm thick quartz radiator in front of the PMT. 
This idea was tested in a set-up with   a somewhat 
faster version (Burle 85001-501 P01) of the same MCP PMT with  10~$\mu$m pores (instead of 
 25~$\mu$m), and with a pulsed laser as a source of photons \cite{vavra-vci07}. 
 In Fig.~\ref{fig:tres-vavra} the improvement in resolution is clearly  observed.
By using an even thicker radiator (up to 5~cm), a time resolution of 6~ps was observed 
\cite{inami-5ps}. With such a resolution 
the time-of-flight counter becomes quite competitive with a RICH counter
in the kinematical region up to ~4~GeV/$c$.

\section{Summary}
\label{summary}

Particle identification devices, in particular the \v Cerenkov counters
 of the BaBar and Belle spectrometers 
have contributed significantly to the rich  $B$ physics harvest of the last seven years.
Identification of particles in the final state with \v Cerenkov counters
has also contributed to the present understanding of nucleon structure and of the 
production of heavy quarks in nucleon-nucleon collisions.
Upgrades are planned to further improve their performance  and to allow them to work at even 
higher event and background rates. New techniques, including a novel focusing radiator concept and 
new fast photon detectors, have been successfully studied. 
By using the time of arrival of \v Cerenkov photons, the dispersion error could be reduced.
Furthermore, a RICH counter equipped with fast photon detectors
 can  be  combined with a time-of-flight or a
time-of-propagation  counter.  Finally, a new type of TOF counter was studied 
with the same very fast photon detectors. 
It is expected that these particle identification methods will again play a decisive 
role in the next generation of precision experiments in $B$ physics.

\end{document}